\documentclass[11pt]{article}
\usepackage[pdftex]{graphicx}
\usepackage[numbers]{natbib}
\usepackage{times}
\usepackage{amsmath, amssymb, amstext}
\usepackage[margin=2cm]{geometry}

\begin{document}

\thispagestyle{empty}
%
\title{SIMULATION OF IMPACT AND FRAGMENTATION WITH THE MATERIAL POINT METHOD}
\author{B. Banerjee, J. E. Guilkey, T. B. Harman, J. A. Schmidt, and
        P. A. McMurtry\\
        Department of Mechanical Engineering, University of Utah, 
        Salt Lake City, UT 84112, USA}
\maketitle
\begin{abstract}
The simulation of high-rate deformation and failure of metals is has traditionally been performed using Lagrangian finite element methods or Eulerian hydrocodes.  Lagrangian mesh-based methods are limited by issues involving mesh entanglement under large deformation and considerable complexity in handling contact.  On the other hand, Eulerian hydrocodes are prone to material diffusion.  In the Material Point Method (MPM), the material state is defined on solid Lagrangian particles.  The particles interact with other particles in the same body, with other solid bodies, or with fluids through a background mesh.  Thus, some of the problems associated with finite element codes and hydrocodes are alleviated.  Another attractive feature of the material point method is the ease with which large deformation, fully coupled,  fluid-structure interaction problems can be handled.  In this work, we present MPM simulations that involve large plastic deformations, contact,  material failure and fragmentation, and fluid-structure interaction.

\hspace{13pt}
The plastic deformation of metals is simulated using a hypoelastic-plastic stress update with radial return that assumes an additive decomposition of the rate of deformation tensor.  The Johnson-Cook model and the Mechanical Threshold Stress model are used to determine the flow stress.  The von Mises and Gurson-Tvergaard-Needleman yield functions are used in conjunction with associated flow rules.  Failure at individual material points is determined using porosity, damage and two bifurcation conditions - the Drucker stability postulate and the acoustic tensor check for loss of hyperbolicity.  Particles are converted into a new material with a different velocity field upon failure.  Impact experiments have been simulated to validate these models using data from high strain rate impact experiments.  Finally, results from simulations of the fragmentation of steel containers due to explosively expanding gases are presented.   The results show that MPM can be used as an alternative method for simulating high strain-rate, large deformation impact, penetration, and fluid-structure interaction problems.
\end{abstract}

\section{INTRODUCTION}
  Dynamic failure of metals has been the focus of considerable experimental 
  investigation (\citet{Curran87} and references therein).  Computational
  modeling and simulation of complex impact, penetration, and fragmentation 
  problems has become possible with the rapid improvement in computational
  tools and power (see \citet{Zukas90} for a survey of tools available in 1990).
  The computational codes used for the simulation of these problems can be
  classified as Eulerian or Lagrangian with advantages and disadvantages
  (\citet{Anderson88}) depending upon the framework used.  Recent
  simulations of impact, ductile failure, and fragmentation have tended to use 
  Lagrangian approaches (\citet{Camacho97,Johnson01}) with 
  special techniques for simulating fracture and failure.  
  
  \hspace{13pt}
  In this work, impact, penetration, and fragmentation of metals is simulated
  using the Material Point Method (MPM)~(\citet{Sulsky94,Sulsky95}). MPM is a 
  particle method for structural mechanics simulations.  In this method, the 
  state variables of the material are described on Lagrangian particles or 
  ``material points''.  In addition, a regular structured grid is used 
  as a computational scratch pad to compute spatial gradients and to solve the 
  governing conservation equations.  The grid is reset at the end
  of each time step so that there is no mesh entanglement.  An explicit 
  time-stepping version of MPM has been used in the simulations of impact,
  penetration, and fragmentation presented in this work.  

\section{APPROACH}\label{sec:algomodels}
  The MPM algorithm used in this work is based on the description of 
  \citet{Sulsky95} with modifications and enhancements including
  modified interpolants (\citet{Bard04}) and frictional contact 
  (\citet{Bard01}).  The computations have been performed using the massively 
  parallel Uintah Computational Framework (UCF)~(\citet{Dav2000}) that uses
  the Common Component Architecture paradigm (\citet{Armstrong99}). 

  \hspace{13pt}
  A hypoelastic-plastic stress update approach (\citet{Zocher00})
  has been used with the assumption that the rate of deformation tensor can 
  be additively decomposed into elastic and plastic parts.  This choice can be 
  justified because of the expectation of relatively small elastic strains 
  for the problems under consideration.  Two plasticity models for flow stress 
  are considered along with a two different yield conditions.  Explicit 
  fracture simulation is computationally expensive and prohibitive for the 
  large simulations under consideration.  We have chosen to use porosity, 
  damage models, and stability criteria for the prediction of failure 
  (at material points) and particle erosion for the simulation of fracture 
  propagation.  

  \hspace{13pt}
  A particle is tagged as ``failed'' when its temperature is greater than the
  melting point of the material at the applied pressure.  An additional
  condition for failure is when the porosity of a particle increases beyond a
  critical limit.  A final condition for failure is when a bifurcation 
  condition such as the Drucker stability postulate is satisfied.  Upon failure,
  a particle is either removed from the computation by setting the stress to
  zero or is converted into a material with a different velocity field 
  which interacts with the remaining particles via contact.  Either approach
  leads to the simulation of a newly created surface.

  \subsection{Models}
  The Cauchy stress in the solid is partitioned into volumetric and deviatoric
  parts.  Only the deviatoric part of stress is used in the plasticity 
  calculations assuming isoschoric plastic behavior.  The hydrostatic pressure 
  is calculated either using the elastic moduli or from a temperature-corrected 
  Mie-Gruneisen type equation of state (\citet{Zocher00}).  The shear modulus
  and melting temperature are pressure and temperature-dependent
  (\citet{Steinberg80}).  Two temperature and strain rate dependent plasticity 
  models have been used - the Johnson-Cook model (\citet{Johnson83}) and the 
  Mechanical Threshold Stress (MTS) model (\citet{Follans88,Goto00}).  
  In addition, two yield criteria have been explored - the von Mises condition
  and the porosity-dependent Gurson-Tvergaard-Needleman (GTN) yield 
  condition (\citet{Gurson77,Tver84}).  An associated flow rule is used to 
  determine the plastic rate parameter in either case.
  The evolution of porosity is calculated as the sum of the rate of growth 
  and the rate of nucleation (\citet{Chu80}).
  Part of the plastic work done is converted into heat and used to update the 
  temperature of a particle (\citet{Borvik01}).  An equation for the dependence 
  of specific heat upon temperature is used when modeling steel. 
  The heat generated at a material point is conducted 
  away at the end of a time step using the heat equation.  
  After the stress state has been determined, a scalar damage parameter is
  updated using either the Johnson-Cook damage model (\citet{Johnson85}). 
  The determination of whether a particle has failed is made on the 
  basis of either or all of the following conditions: 
  (1) the particle temperature exceeds the melting temperature, 
  (2) the TEPLA-F fracture condition~(\citet{Johnson88}) is satisfied, and
  (3) a bifurcation/material stability condition is satisfied.  Two
  stability criteria have been used - the Drucker stability 
  postulate~(\citet{Drucker59}) and the loss of hyperbolicity criterion 
  (using the determinant of the acoustic tensor) 
  (\citet{Rudnicki75,Becker02}).  

\section{VALIDATION}\label{sec:valid}
\begin{figure}[b]
    \begin{minipage}[b]{45mm}
     \centering
     \scalebox{0.35}{\includegraphics{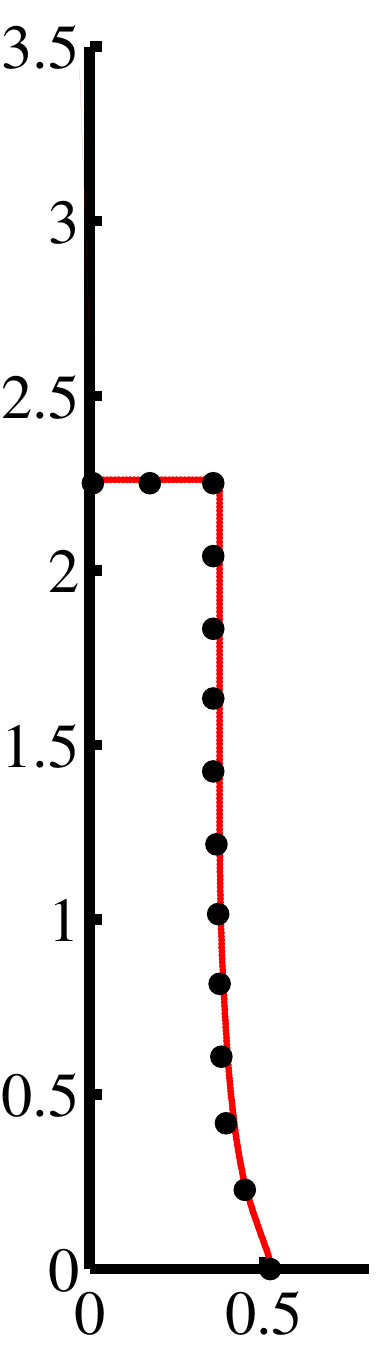}}\\
        (a) 4340 Steel (Johnson-Cook)
    \end{minipage}
    \begin{minipage}[b]{45mm}
     \centering
     \scalebox{0.35}{\includegraphics{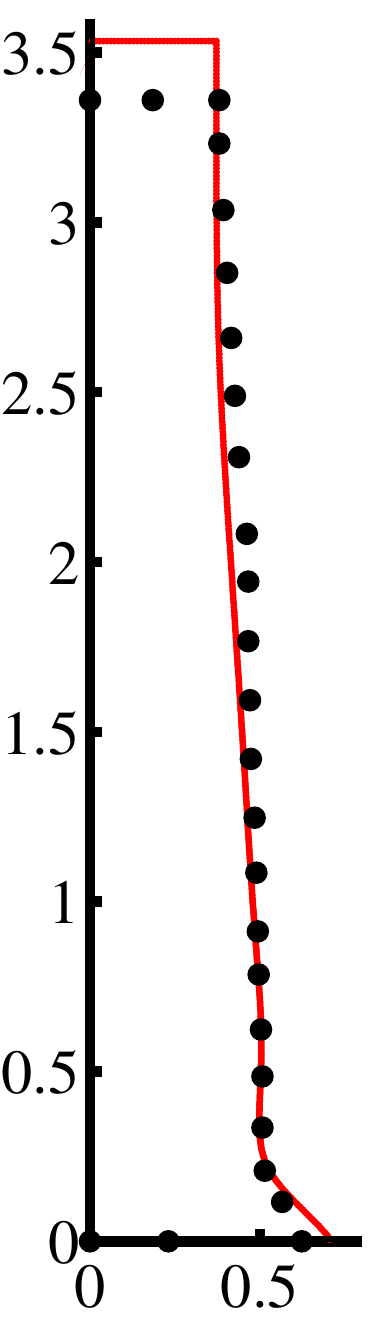}}\\
        (b) Copper  (Johnson-Cook)
    \end{minipage}
    \begin{minipage}[b]{45mm}
     \centering
     \scalebox{0.35}{\includegraphics{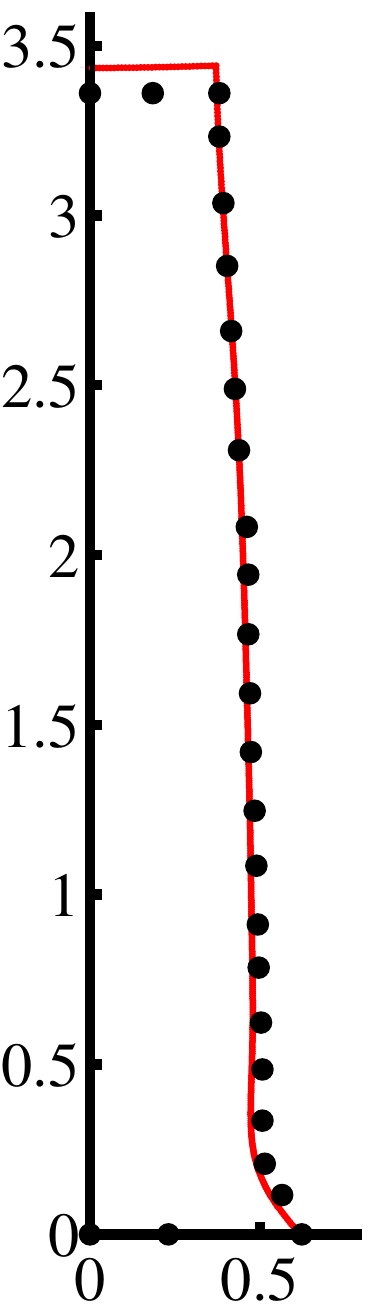}}\\
        (c) Copper (MTS)
    \end{minipage}
    \caption{Simulations of Taylor impact tests (dots = experimental data, solid
             line = simulated data)}
    \label{fig:taylor}
  \end{figure}

  Taylor impact tests have been simulated using MPM to validate the stress
  update procedure and the Johnson-Cook and MTS plasticity models.  
  Figure~\ref{fig:taylor}(a) shows the deformed shape and plastic strain 
  contour ($>$ 0.5) of a 4340 steel cylinder compared with experimental data
  (\citet{Johnson85}).  The simulation results match experimental data 
  remarkably well.
  Figures~\ref{fig:taylor}(b) and (c) compare the simulated deformed shape of 
  an annealed copper cylinder with experimental data (\citet{Zocher00}).
  The Johnson-Cook plasticity model has been used for the result shown in 
  Figure~\ref{fig:taylor}(b) while the MTS model has been used in 
  Figure~\ref{fig:taylor}(c).  A Mie-Gruneisen equation of state has been 
  used in both cases.  The MTS model performs better than the Johnson-Cook 
  model for this material.
  
\begin{figure}[t]
   \begin{minipage}[b]{0.4\textwidth}
     \centering
     \scalebox{0.23}{\includegraphics{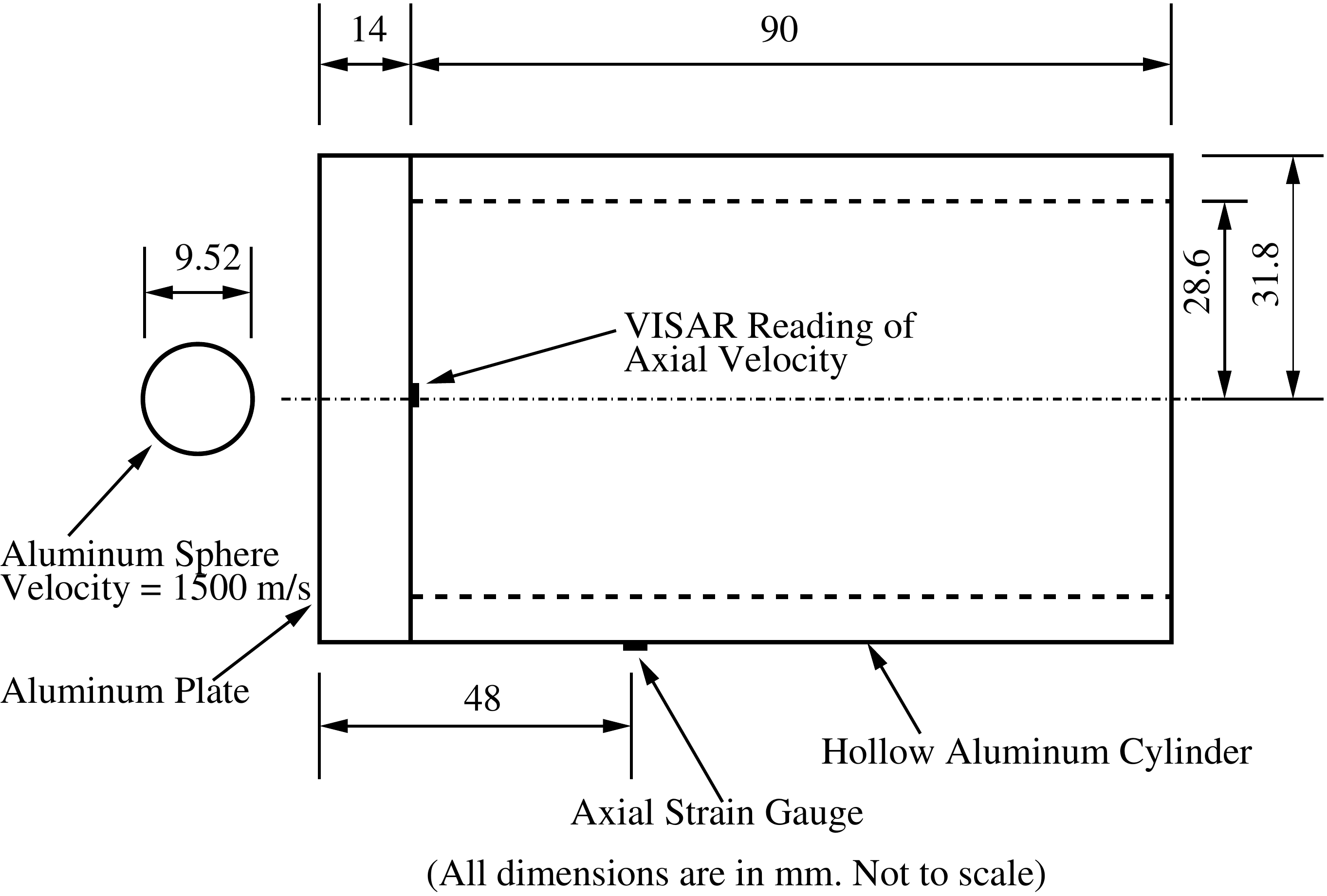}}\\
     (a) Experimental setup. 
   \end{minipage}
   \begin{minipage}[b]{0.6\textwidth}
     \centering
     \scalebox{0.18}{\includegraphics{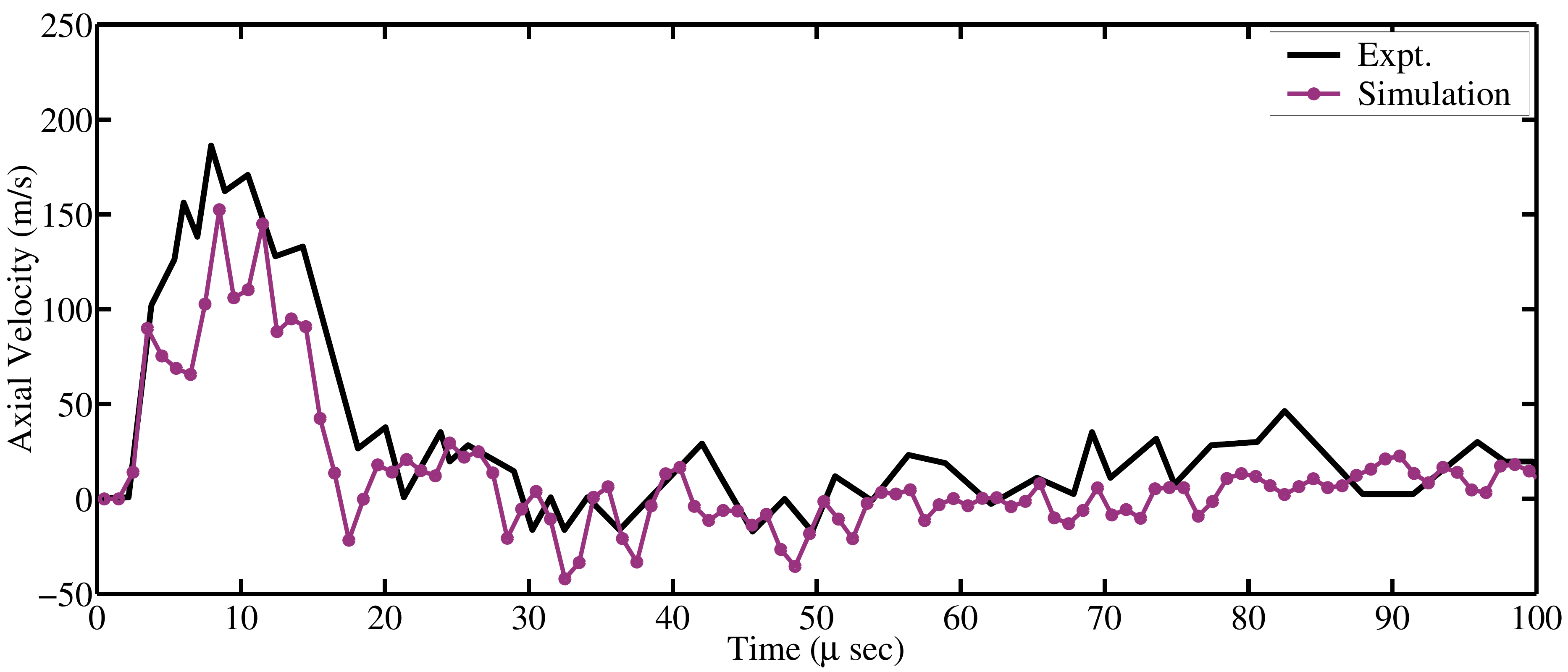}} \\
     (b) Free surface velocity. \\
     \scalebox{0.18}{\includegraphics{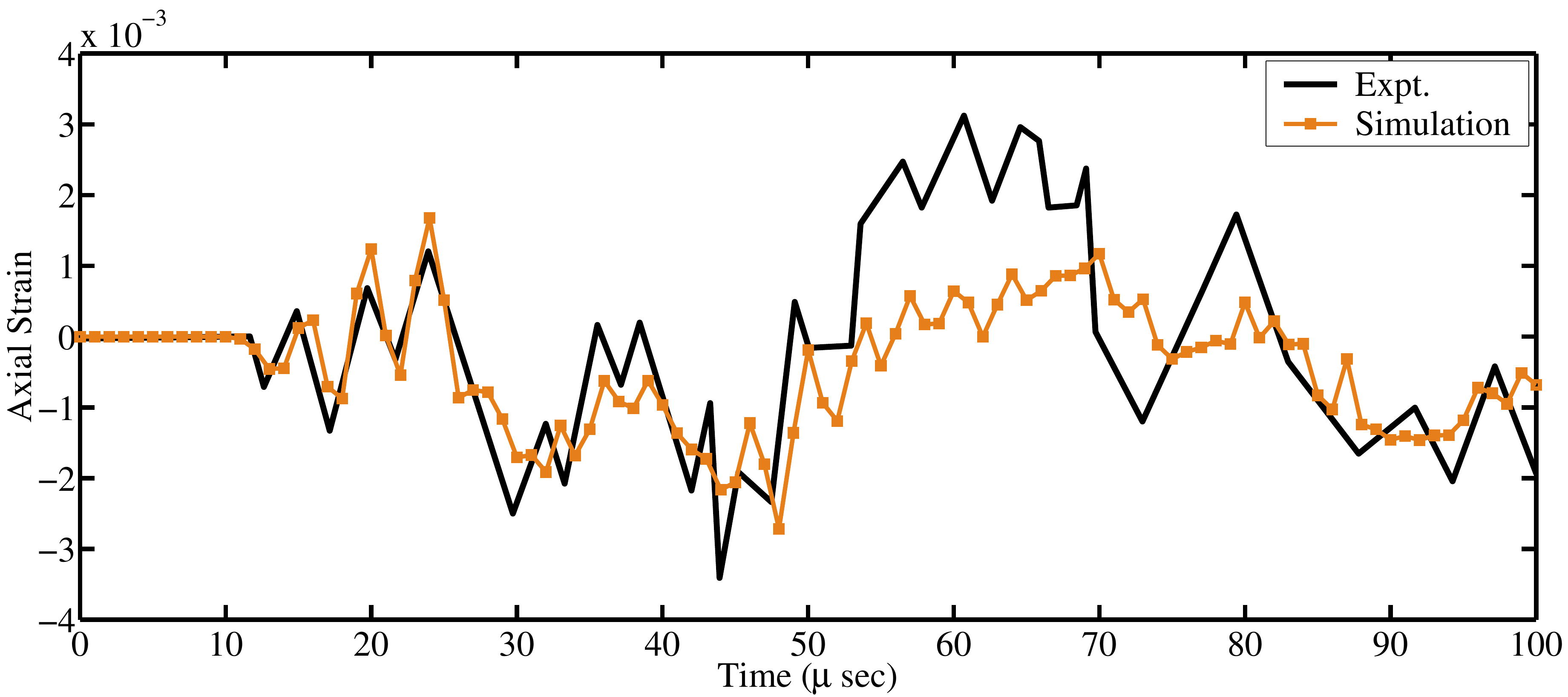}}\\
     (c) Axial strain.
   \end{minipage}
   \caption{Simulations of cylinder impact tests.}
   \label{fig:aluminum}
  \end{figure}

  \hspace{13pt}
  A second validation experiment has been performed by simulating the 
  impact of a 6061-T6 aluminum sphere against a plate attached to a hollow
  cylinder of the same material (\citet{Chhabil99}).  The experimental setup,
  and comparisons of free surface velocity and axial strains are shown in
  Figures~\ref{fig:aluminum}(a), (b), and (c), respectively.
  There is some ringing of the cylinder in the simulations, but the overall
  trend is captured.  Some of the difference between the experimental data
  and the simulations could be because a Johnson-Cook model (\citet{Lesuer01})
  was used for the aluminum.  The above validation tests show that the MPM
  code performs as expected.

\section{SIMULATIONS}\label{sec:simul}
\begin{figure}[t]
   \begin{minipage}[t]{68mm}
     \centering
     \scalebox{0.40}{\includegraphics{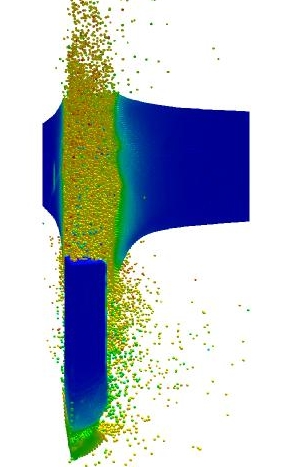}}\\
     (a) Stress set to zero upon failure.
   \end{minipage}
   \begin{minipage}[t]{68mm}
     \centering
     \scalebox{0.40}{\includegraphics{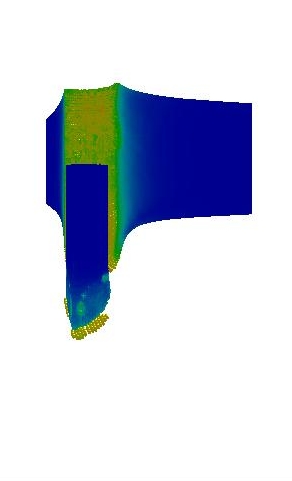}}\\
     (b) Particles converted upon failure.
   \end{minipage}
   \caption{Simulations of penetration (particles colored by plastic strain).}
   \label{fig:pene}
  \end{figure}

  The impact and penetration of a S7 tool steel projectile into an Armco Iron
  target has been simulated using MPM with two different particle erosion
  algorithms.  The geometry of the test is from \citet{Johnson02} and the 
  material properties have been obtained from \citet{Johnson85}.  The depth 
  of penetration after 160 $\mu$s is shown in Figure~\ref{fig:pene}(a) and (b).
  Both cases use frictional contact.  The depth of penetration is less 
  for the case when particles are converted into a new material after failure.
  Also, the energy balance is better behaved in that case.  There is some
  mesh dependence on the depth of penetration which is currently under 
  investigation.

\begin{figure}[t]
    \begin{minipage}[t]{45mm}
    \centering
    \scalebox{0.2}{\includegraphics{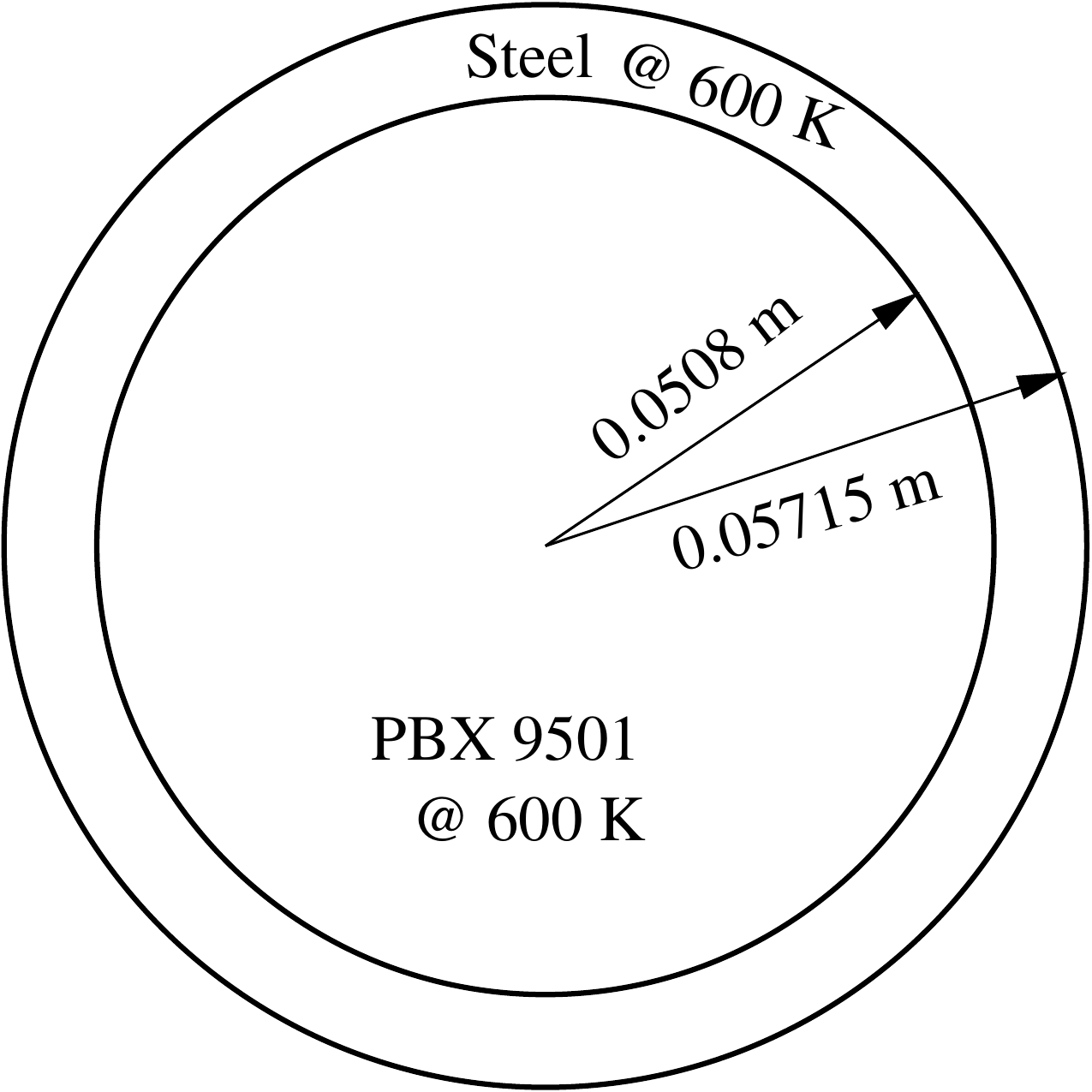}}\\
    (a) Geometry.
    \end{minipage}
    \begin{minipage}[t]{45mm}
    \centering
    \scalebox{0.2}{\includegraphics{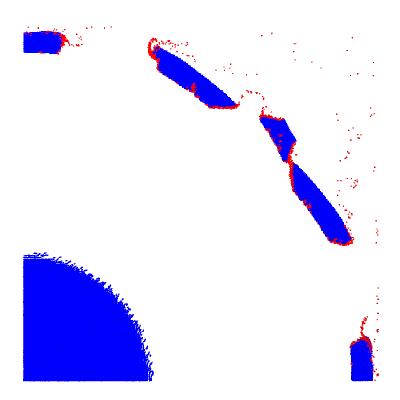}}\\
    (b) 4340 Steel.
    \end{minipage}
    \begin{minipage}[t]{45mm}
    \centering
    \scalebox{0.2}{\includegraphics{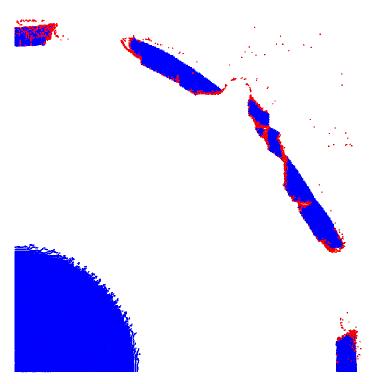}}\\
    (c) HY 100 Steel. 
    \end{minipage}\\
    \begin{minipage}[t]{68mm}
    \centering
    \scalebox{0.7}{\includegraphics{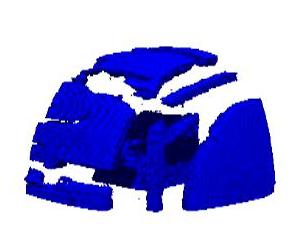}}\\ 
    (d) Fragments of the container. 
    \end{minipage}
    \begin{minipage}[t]{68mm}
    \centering
    \scalebox{0.7}{\includegraphics{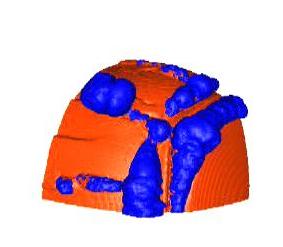}}\\
    (e) Gases escaping from the container.
    \end{minipage}
    \caption{Simulations of fragmenting cylinders.}
    \label{fig:fragments}
  \end{figure}

  \hspace{13pt}
  We have also simulated a coupled fluid-structure interaction problem 
  where a cylinder expands and fragments due to gases generated inside.
  The dynamics of the solid materials - steel and PBX 9501 - is modeled 
  using MPM.  Gas-solid interaction is accomplished using an Implicit 
  Continuous Eulerian (ICE) multi-material hydrodynamic code 
  (\citet{Guilkey04}).  A single computational grid is used for all the 
  materials.  The first set of simulations was performed using the geometry 
  shown in Figure~\ref{fig:fragments}(a).  A steel cylinder was used to confine
  the PBX 9501 material and the simulation was started with both materials at 
  a temperature of 600 K.  An initial Gaussian distribution of porosity
  was assigned to the steel.  The fragments of the cylinder after failure
  (for two steels - 4340 and HY100) are shown in Figures~\ref{fig:fragments}(b)
  and (c).  The Johnson-Cook model was used for 4340 steel.  The MTS model
  (\citet{Goto00}) and the GTN yield condition was used for HY100.
  The expected number of fragments along the circumference matches 
  the analytical prediction by ~\citet{Grady92}.  Both steels show similar 
  fragmentation though the exact shape of the fragments differs slightly.  

  \hspace{13pt}
  Figure~\ref{fig:fragments}(d) and (e) shows the fragmentation obtained from 
  three-dimensional simulations of a 4340 steel cylinder with end-caps
  containing PBX 9501.  The simulation was started with both materials at 
  a temperature of 600 K.  A uniform initial porosity was assigned
  to all steel particles and evolved according to the models discussed in
  the previous section.  Upon failure, the particle stress was set to zero.
  The figures show that these simulations capture some of the qualitative 
  features observed in the experiments on exploding steel cylinders. 

\section{DISCUSSION AND CONCLUSION}\label{sec:concl}
  A computational scheme for the simulation of high rate deformation, impact,
  penetration and and fragmentation using the material point method has
  been presented.  Various impact tests have been used to verify and 
  validate the approach.
  Simulations of target penetration have shown that energy is better conserved
  when particles are converted into materials with a different velocity field
  upon failure (rather than when the stress is set to zero).  Some mesh
  dependence of the results has been observed.
  Simulations of exploding cylinders in two-dimensions have been compared 
  with analytical solutions for the expected number of fragments and found 
  to provide good agreement.  Three-dimensional simulations also
  show qualitative agreement with experiments.   These results show that
  the material point method is an excellent tool for the simulation of 
  high rate deformation and fragmentation problems.

\section*{Acknowledgments}
  This work was supported by the the U.S. Department of Energy through the 
  Center for the Simulation of Accidental Fires and Explosions, under grant 
  W-7405-ENG-48.

\bibliographystyle{unsrtnat}
{\footnotesize
\bibliography{mybiblio}
}
\end{document}